\documentclass[aps,amsmath,reprint,amssymb,groupedaddress]{revtex4-1}
\usepackage{graphicx}
\usepackage{dcolumn}
\usepackage{bm}
\usepackage{bm}
\usepackage{gensymb}%
\usepackage[english]{babel}
\usepackage{txfonts}

\begin{document}

\title{Using deformable particles for single particle measurements of velocity gradient tensors} 

\author{Bardia Hejazi$^1$, Michael Krellenstein$^1$, Greg A. Voth$^1$}
\email{gvoth@wesleyan.edu}
\affiliation{$^1$Department of Physics, Wesleyan University, Middletown, Connecticut 06459, USA}

\date{\today}

\begin{abstract}

We measure the deformation of particles made of several slender arms in a two-dimensional (2D) linear shear and a three-dimensional (3D) turbulent flow. We show how these measurements of arm deformations along with the rotation rate of the particle allow us to extract the velocity gradient tensor of the flow. 
The particles used in the experiments have three symmetric arms in a plane (triads) and are fabricated using 3D printing of a flexible polymeric material.
Deformation measurements of a particle free to rotate about a fixed axis in a 2D simple shear flow are used to validate our model relating particle deformations to the fluid strain.
We then examine deformable particles in a 3D turbulent flow created by a jet array in a vertical water tunnel. Particle orientations and deformations are measured with high precision using four high speed cameras and have an uncertainty on the order of $10^{-4}$ radians.
Measured deformations in 3D turbulence are small and only slightly larger than our orientation measurement uncertainty. 
Simulation results for triads in turbulence show deformations similar to the experimental observations.
Deformable particles offer a promising method for measuring the full local velocity gradient tensor from measurements of a single particle where traditionally a high concentration of tracer particles would be required.

\end{abstract}

\maketitle

\section{Introduction}
\label{intro}

The deformation of structures by fluid flows plays a central role in many environmental, biological, and engineering situations and so has long been a focus of research~\citep{Stockie1998,Alben2002,Qian2008,Ezzeldin2015,Guo2017,Marchioli2019,Rosti2019}.  Recently, the ability to measure and simulate the rotations~\citep{Voth2017} and deformations~\citep{Allende2018,Gay2018,Dotto2019,Rosti2018} of small particles advected in turbulent flows has provided substantial new insights about particle-turbulence interactions as well as the structure of turbulence itself.   But a clear quantification of what features of the flow can be measured from deformations of particles has been lacking.  

Fibers have been a primary focus of prior work on deformable particles in turbulence.  This is natural since slender fibers are easily deformable and simple to fabricate.    However the deformation of fibers is a special case that is somewhat more complex than the deformation of other particles.     A flow with a uniform velocity gradient rotates but does not bend straight material lines.  Fibers can still deform in uniform velocity gradients if there is a compressive stress that exceeds the buckling threshold.  Below this threshold, deformation of fibers is a result of curvature of the velocity field. 

Buckling is caused by a compressive viscous drag acting on a fiber.  For turbulent velocity gradients characterized by shear rate $\dot{\gamma}=(\sqrt{15}\tau_\eta)^{-1}$, the axial viscous force from slender body theory is proportional to $\mu L^2/[\tau_{\eta}\log(2\kappa)]$, where $L$ is the fiber length, $\mu$ is the dynamic viscosity, $\tau_{\eta}$ is the Kolmogorov time scale of the flow, and $\kappa$ is the aspect ratio ($L/D$) where $D$ is the fiber diameter.
Buckling is resisted by the elastic restoring force which scales as $EI/L^2$, where $E$ is the Young's modulus and $I$ is the second moment of area which is proportional to $D^4$ for cylindrical fibers.
Buckling instabilities occur when the ratio of compressive viscous drag to the elastic restoring force $Z=2\pi\mu L^4/[\sqrt{15}\tau_{\eta}EI\log(2\kappa)]$ exceeds a threshold.   Several different conventions for the constant in this non-dimensional number exist in the literature, and we have chosen to follow \citet{Becker2001} who find that for a free fiber in pure strain flows, the buckling instability occurs at $Z\geq153.2$~\citep{Young2007,Becker2001}.  In turbulent flows, much larger values of $Z$ are necessary to obtain buckling in a significant fraction of particles~\citep{Allende2018}.  
\citet{Brouzet2014} experimentally measured deformations of long fibers in turbulence.  For fibers short enough to be in uniform velocity gradients, their approach of quantifying turbulent power produces the same scaling as the buckling instability.  In practice it is quite difficult to obtain particles flexible enough and viscous stresses large enough to observe buckling of fibers in turbulence when the fiber length is small enough to experience uniform velocity gradients.  For example, the particles we use in this paper have $0.005<Z<0.007$, and are still not shorter than the Kolmogorov length.

Initial deformation of a straight fiber at shear rates below the buckling instability comes from curvature of the velocity field.   
Velocity differences over the length of the fiber due to curvature of the velocity field scale as $L^2/(\tau_{\eta} \eta)$, where $\eta$ is the Kolmogorov length scale of the flow, so the viscous drag due to curvature is proportional to $\mu L^3/[\tau_{\eta}\eta\log(2\kappa)]$. 
The ratio of the viscous drag and elastic restoring force in this case is then proportional to $Z L/\eta$.
Thus bending due to curvature is less important than buckling for small fibers with $L/\eta \ll 1$.   However, bending does not have a threshold, so small deformations of fibers near the Kolmogorov length could be dominated by curvature rather than buckling in certain conditions.  
  For longer fibers, curvature and buckling both play a role which leads to fascinating dynamics~\citep{Gay2018,Ali2016,Rosti2018}, but makes it difficult to use deformation as a measurement technique.

However, there exists a simpler regime where particle deformation can be related to local uniform velocity gradients without a buckling threshold.
The key is to leave the slender limit since a particle that is spatial extended in more than one direction will deform in a uniform strain. 
In practice most particles that are not slender fibers are too rigid to deform in the fluid velocity fields typically encountered in turbulence, so this approach may seem impractical.   However, a particle made of several connected slender arms will deform in uniform strain and can be made highly flexible.
In a uniform velocity gradient, a particle made of flexible cylindrical arms will deform by an angle that scales as $\delta \theta \propto \mu L^4 /(\tau_{\eta}EI)\propto Z$, the same scaling as the buckling transition but without a threshold or directional selectivity. 
The effects are even greater if a weak link is introduced so that the arm flexibility is larger than that of a straight cylinder. If an arm bends at its weak link with a spring constant $k$ when torque $\tau$ is applied, $\tau=k\delta\theta$, then the deformation due to uniform velocity gradients scales as 
\begin{equation}
\delta \theta\propto \mu L^3/(k \tau_{\eta}),
\label{deltheta}
\end{equation}
 which is lower order in the aspect ratio of the arm than buckling and bending from curvature in the velocity field. A more detailed derivation of this result is offered in the appendix.

In this paper, we show that particles made of slender arms with weak links near the junction are promising tools for measuring the full velocity gradient tensor in incompressible flow by imaging a single particle trajectory.  Such particles can be fabricated using 3D printing of flexible plastics. The rotation of such particles have been measured in previous work~\citep{Marcus2014}.  If the particle has high enough symmetry and the deformations are small, it rotates just like a sphere~\citep{Bretherton1962,Harris1975} and provides a measurement of the vorticity.   The deformation of the arms from their equilibrium angles provide a measurement of all elements of the strain rate tensor as long as viscosity dominates inertia so that they stay in the quasi-steady, overdamped regime where the elastic restoring force balances viscous drag from the strain rate.

An advantage of using deformable particles is the ability to extract the velocity gradient with only a single particle. Traditionally, to measure the strain and vorticity of the flow we would need to track several particles in a region of space over which the velocity gradient is uniform, which in complex flows occurs at very small length scales.  As a result, a large concentration of tracer particles is required for such measurements.  Often these concentrations become so high that it is difficult to image very far from the surface of the flow.  Making these measurements with deformable particles could allow access to velocity gradients along particle trajectories in locations where they are currently very difficult to measure.
 
To test this idea, we have performed two sets of experiments on the deformations of triads formed from three arms in a plane with weak links near the central axis as shown in Fig.~\ref{triads}.  We first measured the deformation of a triad particle fixed on a low friction axis in a viscous 2D simple shear flow.   Here the particle arm deformations are 0.04 radians and we show how to extract the 2D velocity gradient tensor from measured arm deformations.  We then performed an experiment with similar triad particles in 3D turbulence in a water tunnel with a grid and active jet array.   These triads are larger than the Kolmogorov scale in this flow so they sample a coarse grained velocity gradient.  Also, because they are planar they do not respond to all components of the strain rate tensor.  But they do allow us to demonstrate that the deformations of 3D printed particles in water can be measured.   We show that we can achieve precision on the order of $10^{-4}$ radians in measured arm orientations which allows us to resolve particle deformations. 
We also simulated the dynamics of small triads in turbulent velocity gradients from the Johns Hopkins database and found deformations qualitatively similar to those measured.  

\section{Theory}
\subsection{Model of arm deformations due to viscous flow}
\label{model}

For small deformations that occur slowly compared with the elastic response time of the particle, the arm deforms until the elastic torque balances the torque applied by the fluid.   We will use resistive force theory to determine the fluid torque~\citep{Gray1955}.   We choose the drag coefficient in the resistive force theory so that the fluid force on a fiber in a uniform flow matches Bachelor's slender body theory~\citep{Batchelor1970,Roy2019}, 
\begin{equation}
\mathbf{F}=4 \pi \mu \mathbf{u} L (\log 2\kappa)^{-1} [(\cos \alpha)\hat{\mathbf{p}}-2\hat{\mathbf{e}}_{\mathbf{u}}],
\end{equation}
where $\mathbf{u}$, with unit vector $\hat{\mathbf{e}}_{\mathbf{u}}$, is the relative velocity between the fiber and the fluid. The angle between the arm which is defined by the unit vector $\hat{\mathbf{p}}$ and the fluid velocity is $\alpha$.
Using this force, the torque on a differential length of a slender arm, $dr$, at distance $r$ from the particle center is
\begin{equation}
\label{slender body}
d\mathbf{\tau} = r \hat{\mathbf{p}} \times \frac{8\pi\mu}{\log 2 \kappa}\mathbf{u}dr.
\end{equation}

For a particle advected in a flow with uniform velocity gradients, the relative velocity between the fluid and the arm at a point $r\mathbf{\hat{p}}$ from the center is
$\mathbf{u}=\mathbf{A}r\hat{\mathbf{p}}-\mathbf{\Omega}\times r\hat{\mathbf{p}}$, where $\mathbf{A}$ is the velocity gradient tensor and $\mathbf{\Omega}$ is the solid body rotation rate of the particle. Using this expression for $\mathbf{u}$, the torque from the fluid is found by integrating Eq.~\ref{slender body} along the full length of the arm
\begin{equation}
\label{torque}
\mathbf{\tau} = \frac{8\pi\mu}{\log 2\kappa} \frac{L^3}{3} [\mathbf{\hat{p}} \times (\mathbf{A}\hat{\mathbf{p}}-\mathbf{\Omega}\times \hat{\mathbf{p}})].
\end{equation}

This torque from fluid drag is balanced by a restoring elastic torque which acts to bring the triad arm back to its equilibrium position, 
\begin{equation}
\label{torque balance}
\frac{8\pi\mu}{\log 2\kappa} \frac{L^3}{3} [\mathbf{\hat{p}} \times (\mathbf{A}\hat{\mathbf{p}}-\mathbf{\Omega}\times \hat{\mathbf{p}})] = k (\mathbf{\hat{p}} \times \Delta \mathbf{p}),
\end{equation}
where $k$ is the torsion coefficient and $\Delta \mathbf{p}$ is the small displacement of the arm from its equilibrium (undeformed) position, $\mathbf{ \hat{p}'}$, to its deformed position, $\mathbf{\hat{p}}$.

In the special case of an isotropic particle such as a tetrad, which is comprised of four cylindrical arms symmetrically oriented with respect to each other, the particle rotates with the same rotation rate as the fluid vorticity since the effective ellipsoid of a tetrad is a sphere.
And so the relative velocity between the fluid and an arm of the particle is only due to the strain component of the velocity gradient tensor, reducing the relative velocity to $\mathbf{u}=\mathbf{S}r\hat{\mathbf{p}}$. As such Eq.~\ref{torque balance} simplifies to
\begin{equation}
\label{torque balance isotropic}
\frac{8\pi\mu}{\log 2\kappa} \frac{L^3}{3} (\mathbf{\hat{p}} \times \mathbf{S}\hat{\mathbf{p}}) = k (\mathbf{\hat{p}} \times \Delta \mathbf{p}).
\end{equation}

\begin{figure}
\centering
 \includegraphics[width=0.4\textwidth]{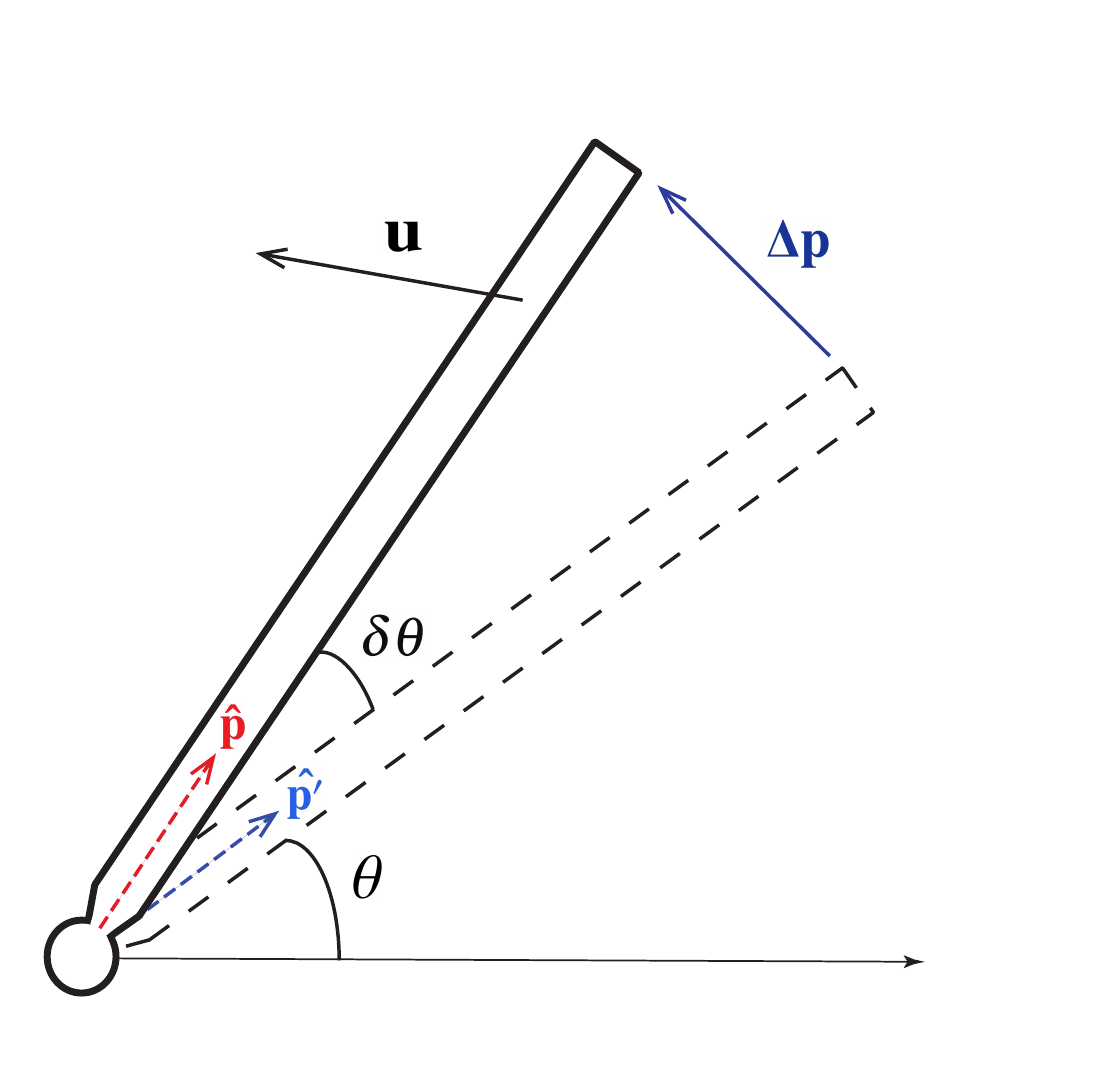}
\caption{A triad arm in its equilibrium position, $\mathbf{\hat{p}}'$, displaced by $\Delta \mathbf{p}$ to its deformed position, $\mathbf{\hat{p}}$, where the perpendicular component of the fluid velocity, $\mathbf{u}$, applies a drag torque on the arm.}
\label{arm sketch}
\end{figure}

Now considering a 2D flow, where all arms of the particle are in the plane, the orientation of arm $n$ is $\theta_n + \delta \theta_n$, where $\theta_n$ is the angle between the undeformed arm and the $x$ axis (see Fig.~\ref{arm sketch}), and $\delta \theta_n=(\hat{\mathbf{p}} \times \Delta \mathbf{p}) \cdot \hat{z}$ is the deformation angle which is assumed to be small. 
For a simple shear flow with $u_{x}=\frac{K}{2}y$ and $u_{y}=\frac{K}{2}x$, where the particle rotates with the fluid and using Eq.~\ref{torque balance isotropic}, the deformations in our 2D flow are
\begin{equation}
\label{2D equation}
\frac{8\pi\mu}{\log 2\kappa} \frac{L^3}{3} |\mathbf{S}| \cos (2(\theta_n+\delta \theta_n)) = k \delta \theta_n,
\end{equation} 
where $|\mathbf{S}|=K/2$.

\subsection{Extracting the strain rate from arm deformations}

We want to extract the 8 free parameters in a three-dimensional incompressible velocity gradient tensor from measurements of the rotation and deformation of a particle.  If a particle is sufficiently isotropic, it rotates with the fluid vorticity so the measured solid body rotation allows reconstruction of the 3  parameters which are the components of the vorticity vector. 
From the measured deformation of the arms, we want to extract the remaining 5 parameters that specify the strain rate tensor.  
We can measure the two angles that specify the orientation of each of the $n$ arms of a particle providing $2n$ measurements.    There are 3 Euler angles needed to specify the orientation of an undeformed particle in space, so the number of arms necessary to determine the 5 remaining parameters is $2n-3 \geq 5$.   Thus we need at least $n=4$ arms whose deformation from their equilibrium orientation is measured in order to measure the strain rate tensor in a 3D flow.     In a symmetric undeformed configuration, these 4 arms would form a tetrad.  

In Eq.~\ref{torque balance isotropic}, there are three equations for each arm in Cartesian coordinates, and so measurements of the orientations of four arms produces 12 equations with degeneracies.
To demonstrate that extracting the strain rate from arm deformations is possible from these equations, we modeled the deformation of a four armed tetrad particle using Eq.~\ref{torque balance isotropic} with a chosen strain rate tensor.  The modeled deformation was then input into Eq.~\ref{torque balance isotropic} and we found that singular value decomposition (SVD) returns a unique solution recovering the correct fluid strain rate in 3D flows.

It would appear that by increasing the number of arms of a system we would have an overdetermined system that would provide a more accurate way of determining the strain rate from measured arm deformations with noise.  However, a 6 armed particle like a jack is a special case that does not work.  
Even though a jack is symmetric and rotates with the flow vorticity, it has co-linear arms which in essence turns the arms on the same axis into a single arm. As a result, the jack will effectively have only 3 arms, which is not sufficient to determine the 3D strain rate tensor.  Other symmetric configurations of arms that have at least 4 non-co-linear arms would work. 

For our experiments in 2D and 3D flows we use triads. The triads have three cylindrical arms in one plane connected to each other at the center of the particle.  The arms are separated equally from one another by ${2\pi}/{3}$ radians. In a 2D simple shear flow using Eq.~\ref{2D equation} we are able to find the full velocity gradient tensor using a triad.   We chose to use triads in our 3D experiments as an initial step to validate our experimental techniques even though they will not be able to measure the full velocity gradient tensor.   The reason for this choice is that they are planar so they are much easier to fabricate on 3D printers, and they can rest on flat surfaces without the highly flexible particles acquiring additional permanent deformations.  

\section{Two-dimensional experiments}
\label{2D}
\subsection{2D Experimental setup}
\label{2D setup}

\begin{figure}
\centering
 \includegraphics[width=0.5\textwidth]{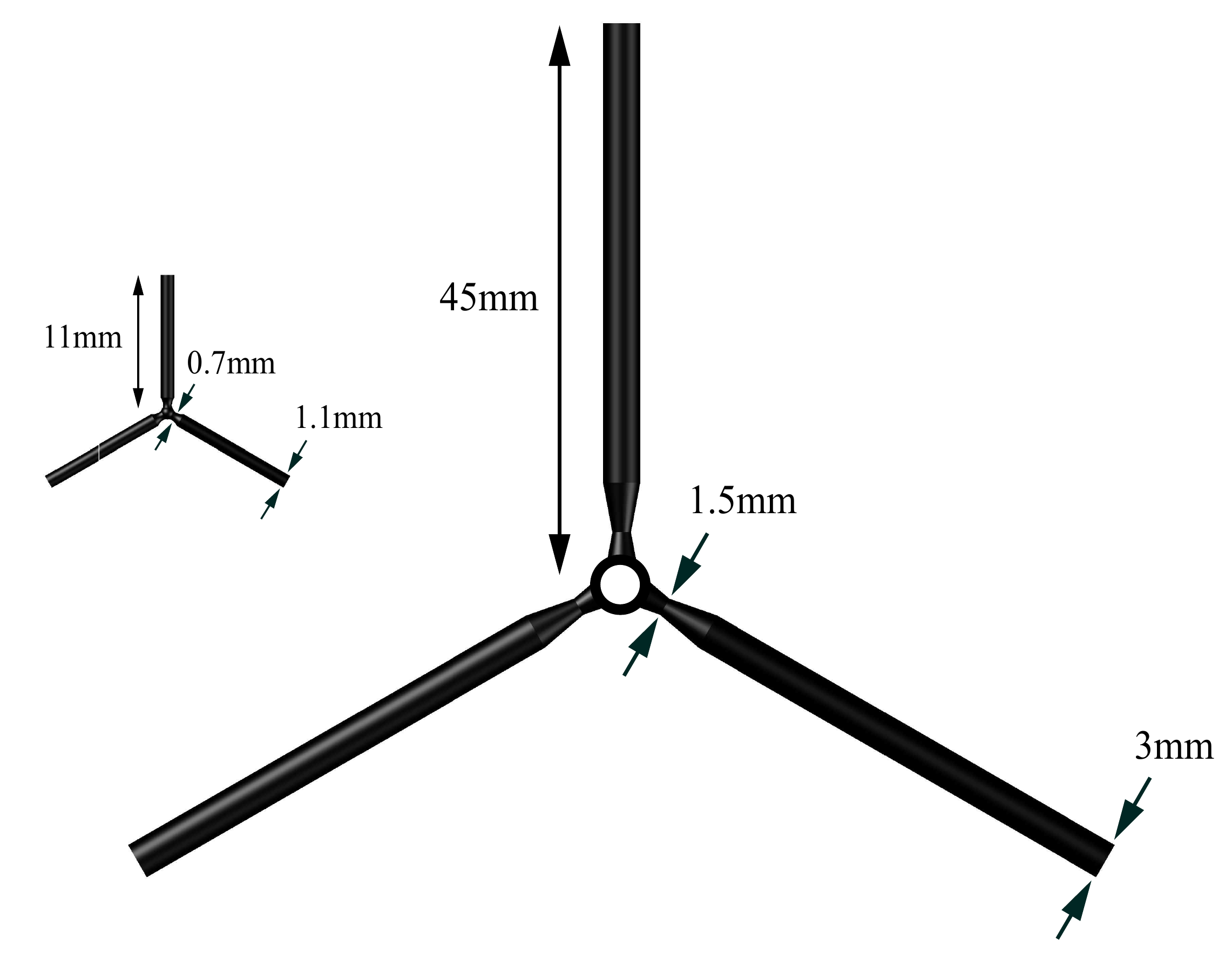}
\caption{The deformable triads used in our experiments. The larger triad with arm length of $L=45$ mm and arm diameter of $D=3$ mm and joint diameter of $1.5$ mm was used in the 2D experiments of simple shear flow. And the smaller triad with arm length of $L=11$ mm and arm diameter of $D=1.1$ mm and joint diameter of $0.7$ mm was used in the 3D turbulence experiments. (The particles are to scale)}
\label{triads}
\end{figure}

Figure~\ref{triads} shows the design for the deformable triads used in our experiments.
The deformable triads are 3D printed with a Formlabs printer using their Elastic Resin which has a Shore durometer of 50A.
Each triad arm tapers to a smaller diameter at a joint near the base of the particle. 
We model the joints as ideal springs, with
a linear relationship between torque and arm deformation angle. The joints also have the advantage of localizing particle deformation making image analysis simpler.  The triad used in the 2D experiments in a simple shear flow has arms with $L=45$ mm and $D=3$ mm that go down to either $1.5$ mm or $1.0$ mm in diameter at the joint.

\begin{figure}
\centering
 \includegraphics[width=0.5\textwidth]{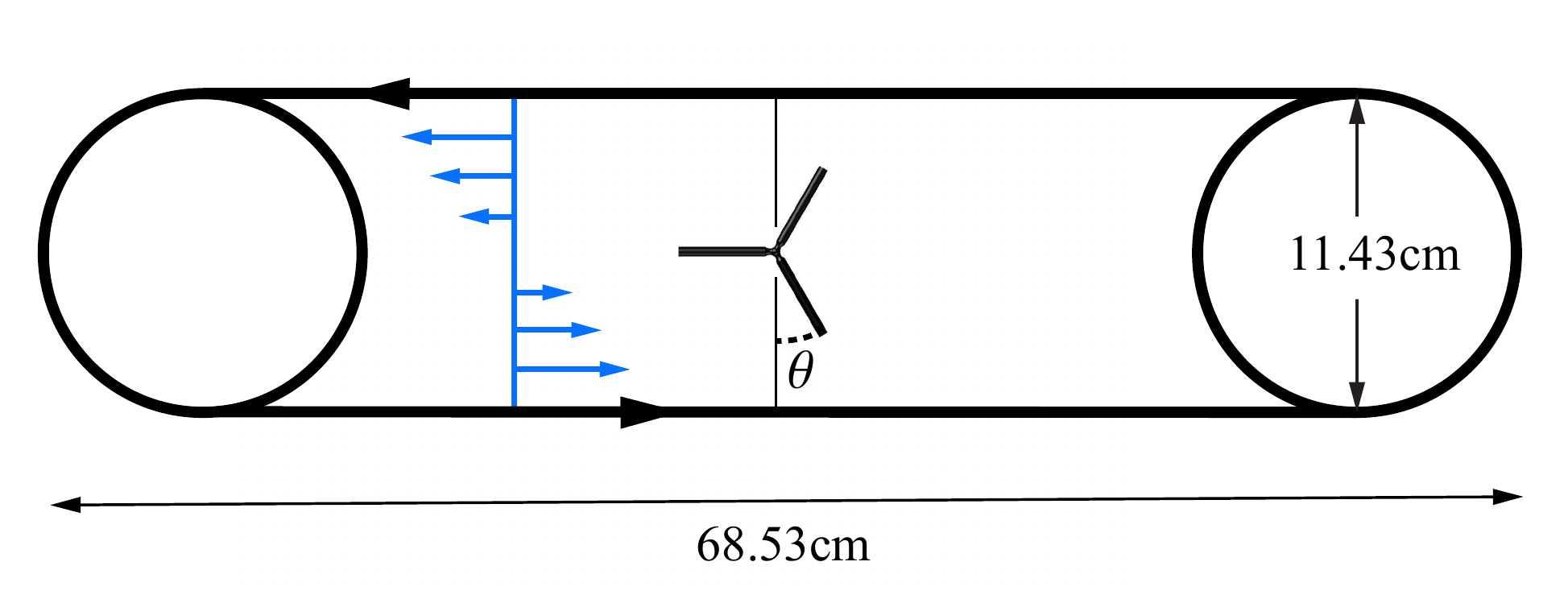}
\caption{Experimental apparatus for the 2D experiments with a simple shear flow created by a moving belt. The particle is suspended in the center of the tank using an air bearing so the particle can rotate at the same rate as the fluid vorticity.}
\label{setup}
\end{figure}

\begin{figure}
\centering
 \includegraphics[width=0.4\textwidth]{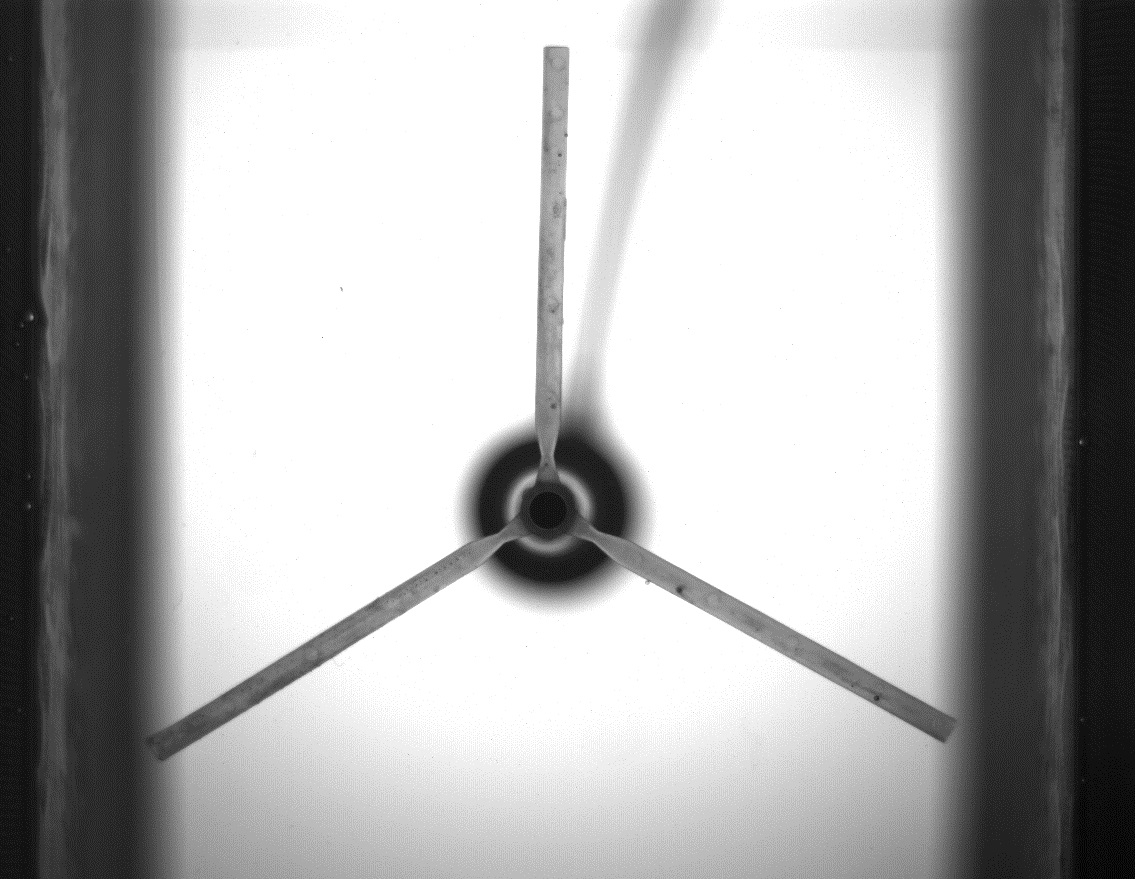}
\caption{Image of a triad in the 2D simple shear flow looking up from the bottom of the flow. The grey areas at the right and left are the belt.  The dark center is the shadow of the air bearing and the shaft that holds the particle.  The curved grey shadow just to the right of vertical is the shadow from the air supply tube.}
\label{2Dimage}
\end{figure}

Simple shear flow is produced by a moving belt as shown in Fig.~\ref{setup}.
The particle is mounted on an air bearing with an axle that extends into the center of the flow. The air bearing allows the particle to freely rotate with the same rate as the fluid vorticity while keeping the particle in a fixed position. As long as the particle is centered where the fluid velocity is zero, fixing its location will not affect its behavior. 
The fluid used is an aqueous glycerol solution of $90$ percent glycerol by volume, with a measured kinematic viscosity of $193 \;\textrm{mm}^2/\textrm{s}$ at a temperature of $19^{\circ}$C. The density of the solution is $1.2 \times 10^{3}\; \textrm{kg}/\textrm{m}^3$~\citep{Volk2018,Segur1951}. The belt is driven by an electric motor which allows us to set the strain rate.
We use backlighting and a 1 megapixel high speed camera located underneath the glass tank to obtain images like the one shown in Fig.~\ref{2Dimage}.
The angles of the arms are extracted by performing non-linear fitting of a Gaussian rod model to the measured images~\citep{Cole2016}.

\begin{figure}
\centering
 \includegraphics[width=0.5\textwidth]{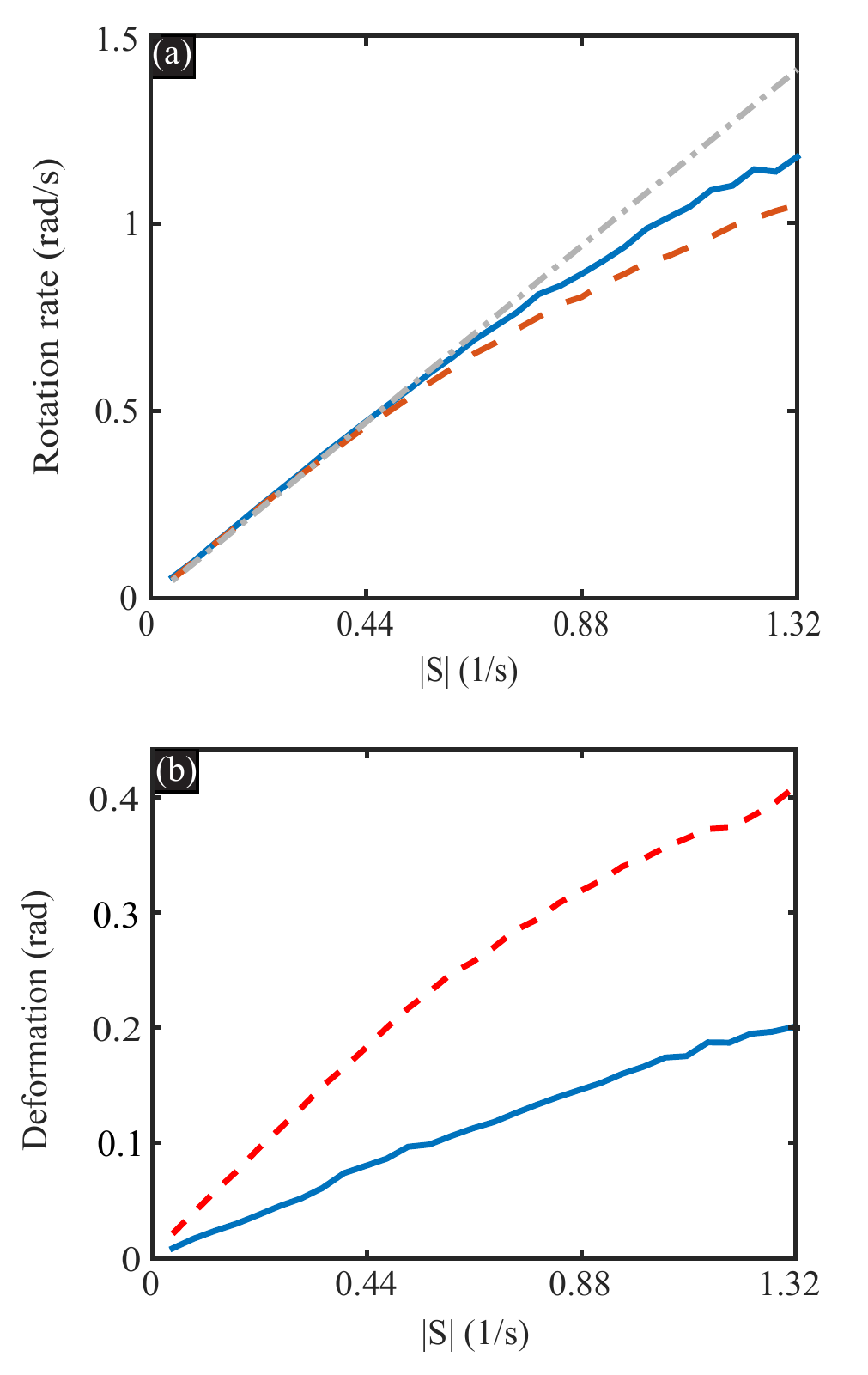}
\caption{(a) The rotation rate  and (b) deformation of a triad particle in 2D simple shear flow as a function of fluid strain rate. The red dashed line is for the particle with joint diameter of $1$ mm and the blue solid line is for the particle with joint diameter of $1.5$ mm. The gray dash-dot line in (a) is the rotation rate for an ideal non deformable particle that always rotates with the same rate as the fluid vorticity. In (b) the deformation is defined as half the peak to peak value of the relative angle between adjacent arms.}
\label{exp paras}
\end{figure}

We initially examined the deformation of two triads with different joint diameters of $1$ mm and $1.5$ mm while keeping all the other parameters the same ($L=45$ mm and $D=3$ mm).
The particle rotation rates and deformations were measured for both particle joint diameters at different strain rates of the flow. Figure~\ref{exp paras}a shows the particle rotation rate as a function of fluid strain rate and Fig.~\ref{exp paras}b shows the 
maximum deformation of the arms as a function of fluid strain rate. The relative deformation between two adjacent arms is sinusoidal in time and here we show the maximum relative deformation calculated by taking half the peak to peak value of the relative arm angles.
From Fig. \ref{exp paras}a we see that at low strain rates both particles experience deformations which are small enough that the particles rotate at the same rate as the fluid. 
But as the strain rate increases and the particles are more deformed, they deviate from being an ideal triad and their effective ellipsoid is no longer a circle.  As a result, the particles tumble in Jeffery orbits with an average rotation rate that is smaller than the fluid. 
The  $1.5$ mm jointed particle is less deformed than the $1$ mm jointed particle at higher strain rates, and so the $1.5$ mm jointed particle behaves like an ideal triad for higher strain rates of the flow.
Furthermore, Fig. \ref{exp paras}b shows that the deformation of both particles has a nearly linear relationship with the fluid strain as predicted by Eq.~\ref{2D equation} up to relative deformations on the order of 0.1 radians. 
 All further experiments in 2D flow were done using the triad with $1.5$ mm joints at a  strain rate of $|\mathbf{S}|=0.35 \;\textrm{s}^{-1}$ where the maximum relative deformation is about 0.07 radians.

\subsection{2D Results}
\label{2D results}

\begin{figure}
\centering
 \includegraphics[width=0.5\textwidth]{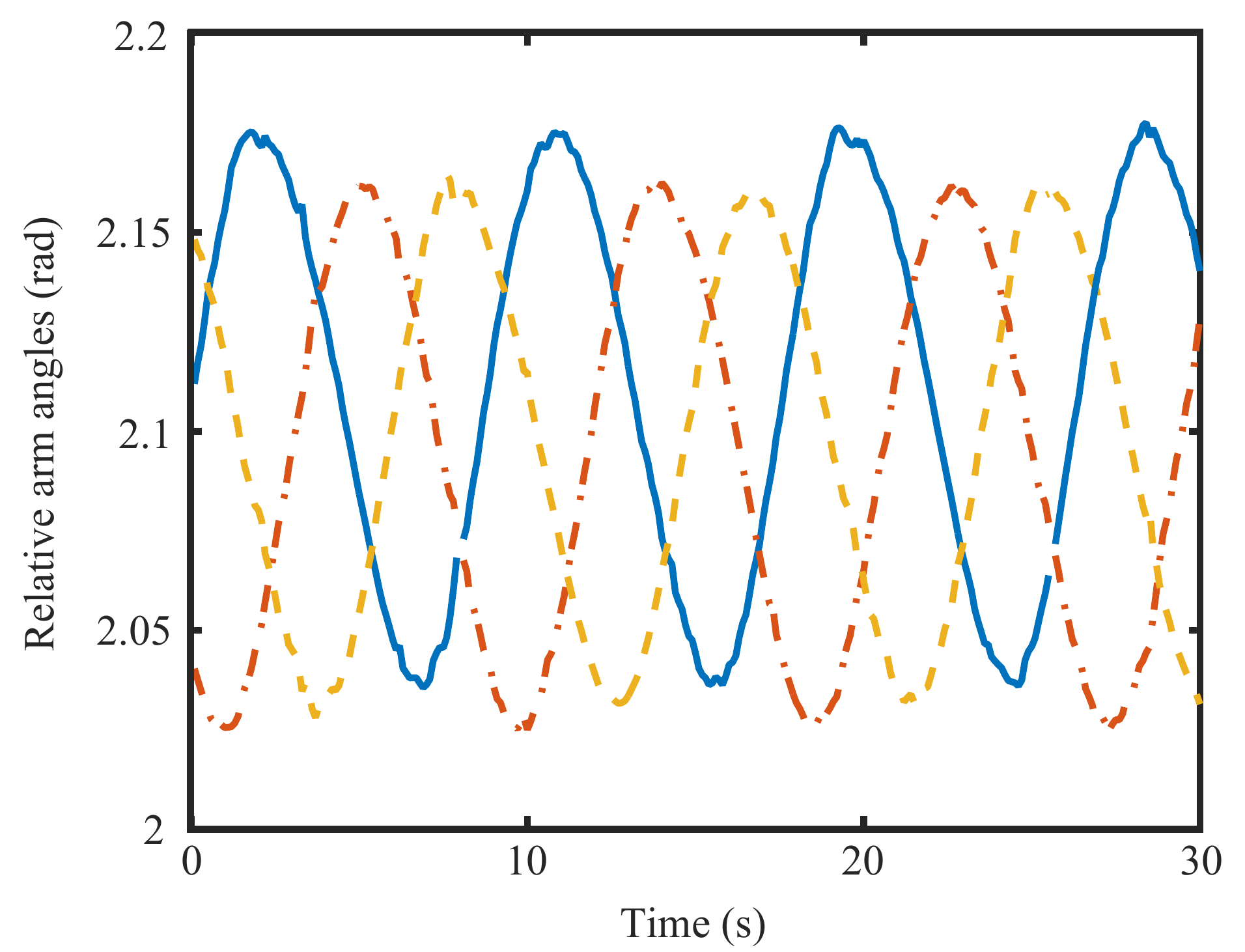}
\caption{Angle between every two neighboring arms as a function of time. For a strain value of $|\mathbf{S}|=0.35 \;\textrm{s}^{-1}$ the angle differences between adjacent arms changes by $\pm 0.07$ radians. Each arm experiences a different amount of deformation depending on the particle orientation at a particular time in the flow. Blue solid line is $\Phi_2-\Phi_1$, red dot-dashed line is $\Phi_3-\Phi_2$, and yellow dashed line is $\Phi_1-\Phi_3$.}
\label{arm defs}
\end{figure}

Figure \ref{arm defs} shows the measured angles between pairs of arms as a function of time.  The arm separation between arms in an ideal triad is $2 \pi /3$ and so the angle differences fluctuate around this value.  The angle difference changes by only $\pm 0.07$ radians which means individual arms separated by $ 2\pi/3$ are deforming by $\pm 0.07/\sqrt{3}=0.04$ radians indicating the precision required to measure this velocity gradient of $|S|$=0.35 s$^{-1}$ with this particle.

\begin{figure}
\centering
 \includegraphics[width=0.5\textwidth]{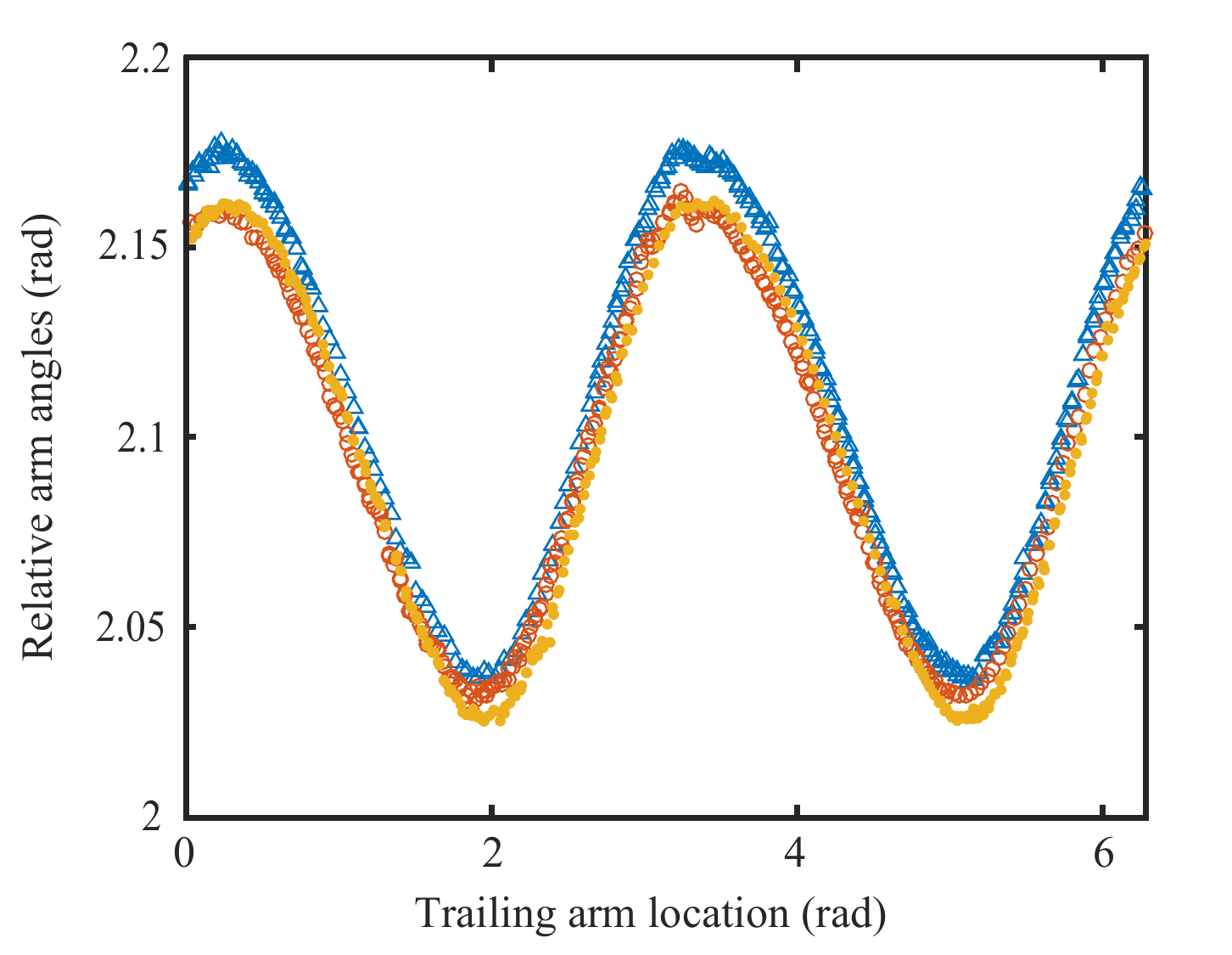}
\caption{The measured angle between a given arm and the arm immediately trailing it as a function of the trailing arms location. Each arm has a unique torsion coefficient that makes the deformation experienced by each arm different. Blue $\bigtriangleup$ shows $\Phi_2-\Phi_1$, red $\circ$ shows $\Phi_3-\Phi_2$, and yellow $\bullet$ shows $\Phi_1-\Phi_3$.($|\mathbf{S}|=0.35 \;\textrm{s}^{-1}$)}
\label{def trailing arm}
\end{figure}

The angle differences between different pairs of arms are slightly different.  This is a result of manufacturing non-idealities that make the equilibrium angles and the torsion coefficients of different arms to be slightly different.  
Figure~\ref{def trailing arm} more clearly reveals the differences between different arms by plotting the angle difference as a function of the angle of the trailing arm. Because of these deviations from ideal particles, we calibrate the particle to quantify the variations in equilibrium arm orientation and torsion coefficients.  

To calibrate our particle, we extend the theory in section~\ref{model} to include deviations from ideal equilibrium arm orientations, $\Delta_n$, and torsion coefficients, $k_n$, that are different for each arm.   For our non-ideal triad, the instantaneous orientations of the arms indexed by arm number $n=1,2,3$ are 
$$ \Phi_n=\theta_1+2\pi (n-1)/3+\Delta_n+\delta \theta_n$$
where arm 1 has been chosen as the reference arm so $\Delta_1$ is defined to be zero.  Then the expression for the deformation of arm $n$ of a real triad is obtained from Eq.~\ref{2D equation},
\begin{equation}
\label{three_2D_equations}
\frac{8\pi\mu}{\log 2\kappa} \frac{L^3}{3} |\mathbf{S}| \cos (2\Phi_n) = k_n \delta \theta_n.
\end{equation}
and the angle between arm one and two is 
\begin{eqnarray}
\label{2D angle diffs}
\Phi_2-\Phi_1&=&\frac{2\pi}{3} + \Delta_2 \; \; \; ... \\
&&
+ \frac{8\pi\mu}{\log 2\kappa} \frac{L^3}{3} |\mathbf{S}|\left[ \frac{1}{k_2} \cos 2\Phi_2 -  \frac{1}{k_1} \cos 2\Phi_1\right] \nonumber
\end{eqnarray}
with similar expressions for $\Phi_3-\Phi_2$ and $\Phi_1-\Phi_3$.  

\begin{figure}
\centering
 \includegraphics[width=0.5\textwidth]{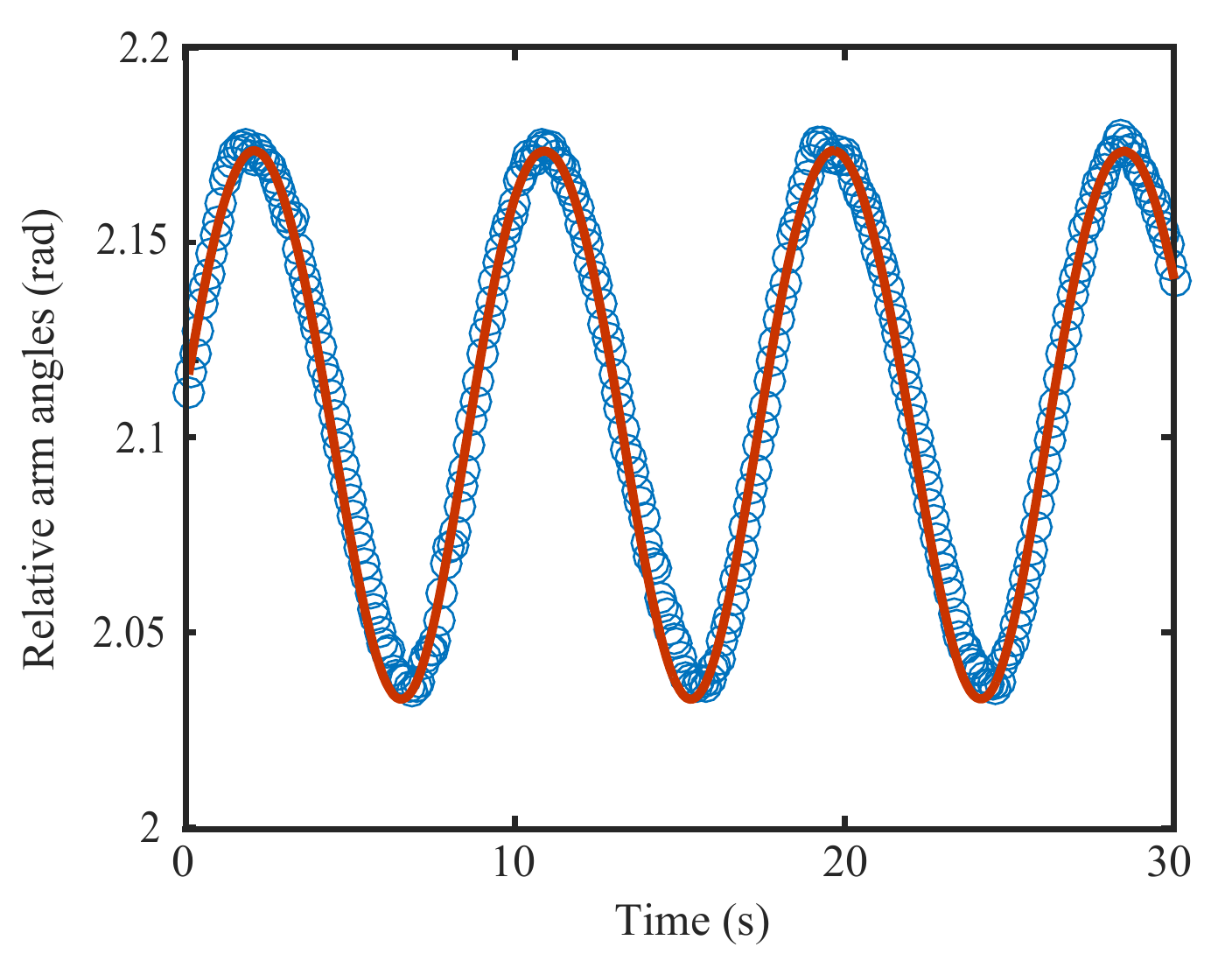}
\caption{Angle between arms 2 and 1, $\Phi_2-\Phi_1$, as a function of time. The solid red line is the prediction from the model in Eq. \ref{2D angle diffs}, and the blue circles are measured angles from experiments. ($|\mathbf{S}|=0.35 \textrm{s}^{-1}$)}
\label{exp model}
\end{figure}

A non-linear fit of the measured angle differences between all the arms to the model in Eq~\ref{2D angle diffs} gives best fit parameters $(k_1, k_2, k_3)=(0.27, 0,25, 0.33)\times 10^{-3}$ N\,m/rad, and \\
$(\Delta_1, \Delta_2, \Delta_3)=(0,0.005,0.002)$ radians. 
 Figure~\ref{exp model} shows the fit of the measured arm angle differences between arms 1 and 2. The other angle differences have a similar match to the fit and are not shown.    The deviations from an ideal triad are small relative to the total deformation of the particle, but we find that correcting for them is essential to obtain accurate measurements of the instantaneous velocity gradients.

\begin{figure}
\centering
 \includegraphics[width=0.5\textwidth]{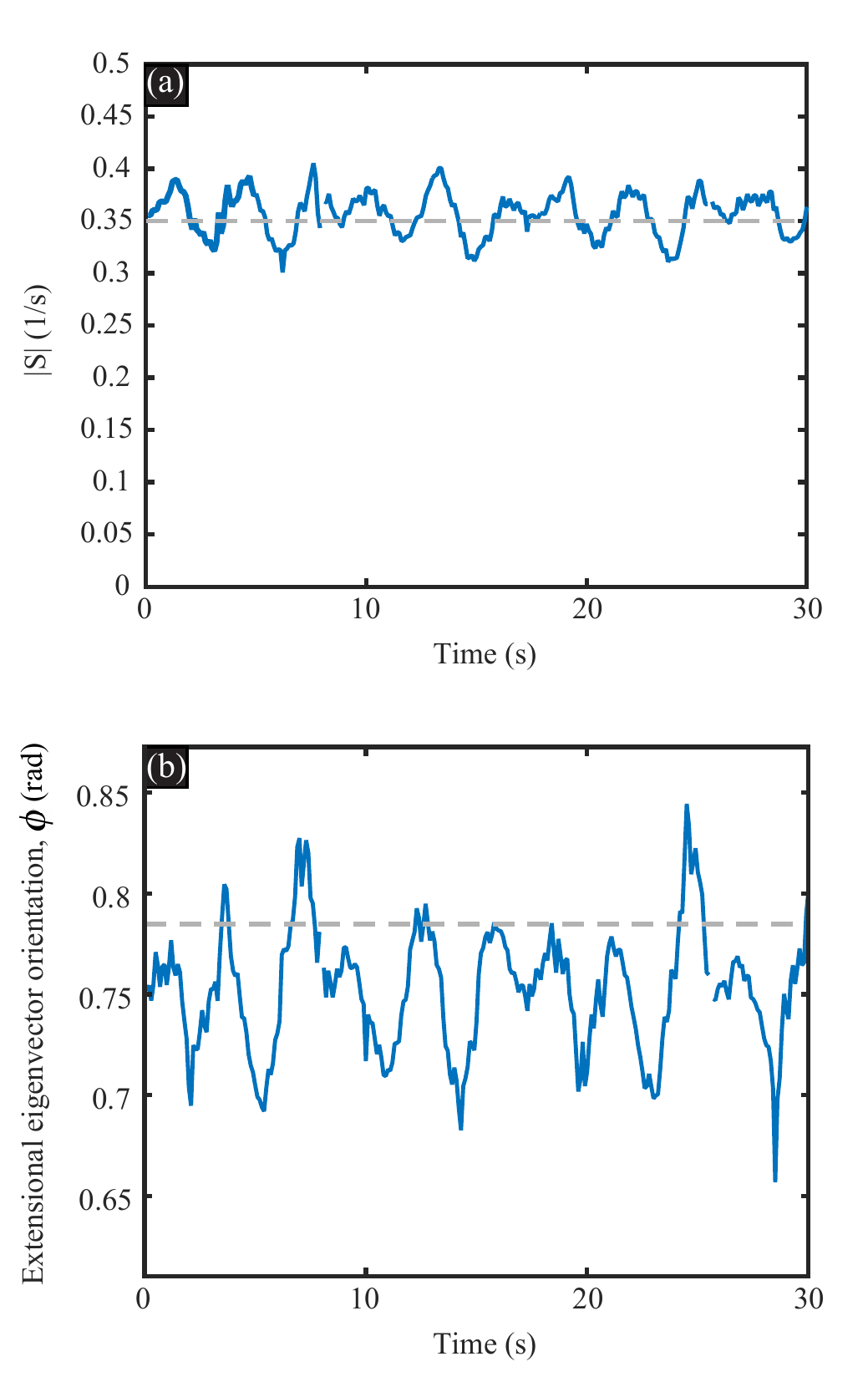}
\caption{(a) Magnitude of the fluid strain where the solid blue line is the calculated value from the model of Eq. \ref{2D equation} and the dashed gray line is the known value of $|\mathbf{S}|=0.35 \textrm{s}^{-1}$. (b) The orientation of the extensional eigenvector of the fluid strain calculated from Eq. \ref{2D equation}. We expect the extensional eigenvector to have an orientation of $\pi/4$ in a simple shear flow as shown by the dashed gray line.}
\label{strain}
\end{figure}

With the particle calibration determined, we are ready to fit for the instantaneous strain rate from measured particle deformations.   Here we add an additional parameter $\phi$ to Eq.~\ref{2D angle diffs} quantifying the angle of the extensional strain eigenvector with respect to its direction in simple shear.   The expression
\begin{eqnarray}
\label{2D angle diffs_with_strain_orientation}
\Phi_2&-&\Phi_1=\frac{2\pi}{3} + \Delta_2 \; \; \; ... \\
&&
+ \frac{8\pi\mu}{\log 2\kappa} \frac{L^3}{3} |\mathbf{S}|\left[ \frac{1}{k_2} \cos (2\Phi_2+\phi) -  \frac{1}{k_1} \cos (2\Phi_1+\phi)\right] \nonumber
\end{eqnarray}
and the similar expressions for $\Phi_3-\Phi_2$ and $\Phi_1-\Phi_3$ can be fit to the instantaneous measurements of $\Phi_1$, $\Phi_2$, and $\Phi_3$ to obtain $|\mathbf{S}|$ and $\phi$.   
Figure~\ref{strain}(a) shows the measured magnitude of the strain rate deduced from instantaneous particle deformation.  The average value matches the actual strain because the particle was calibrated in this flow.  
The new result here is the demonstration that measurements of the deformation of arms of a particle provides a measurement of the instantaneous strain rate tensor in the flow as long as the particle deformation relaxation time scale is shorter than the time scale associated with the velocity gradients.  Figure~\ref{strain}(b) shows the orientation of the strain rate eigenvectors determined from the particle deformation.    The measured orientation is systematically below the expected value of $\pi/4$ by about 0.035 radians or 2.0 degrees, reflecting limitations of alignment of the camera with the simple shear flow.  It would be possible to add this as an additional parameter to the calibration and potentially improve these measurements.

We also performed an independent but crude calibration of the torsion coefficients of the triad used in the 2D experiments.  
This was done by hanging weights near the end of an arm and recording the deformation.
With this method the torsion coefficient was measured to be $k=6.4\times 10^{-4}$ N\,m/rad.  This is a factor of 2 larger than the calibrated values of $k_n$ found from the measured deformation in simple shear flow.  This discrepancy is larger than we expected, but may be a consequence of stiffening of the flexible polymer as a result of age and drying.    Future use of these methods would likely need to either obtain calibrations in situ or check for changing stiffness of the particles over time.

\section{Three-dimensional experiments}
\label{3D}
\subsection{3D Experimental setup}
\label{3D setup}

The 3D experiments on deformable particles were done in a vertical water tunnel with a test section measuring $30 \times 30 \times 150~\textrm{cm}^{3}$. 
The vertical water tunnel creates controlled  turbulence using a random jet-array that creates approximately homogeneous-isotropic turbulence in the test section.  
In these experiments we have high turbulence intensities with $Re_{\lambda}=198 \pm 4$ and a mean fluid velocity of $\langle\mathbf{U}_f\rangle_z=21.51 \pm 3$ mm/s. The parameters of the turbulence in our experiments have been interpolated from data obtained from previous measurements of the water tunnel~\citep{Kramel2018}.
The energy dissipation rate of our tunnel for water at room temperature is $108 \pm 60 \;(\textrm{mm}^2/\textrm{s}^3)$. The energy dissipation rate is obtained from interpolating measurements that use the third order structure functions of the mean fluctuating velocity.  
The tunnel is designed so that the through flow counteracts the sedimentation velocity of particles allowing for particles to stay a long time in a fixed detection volume with constant turbulence statistics.   The large uncertainty in the energy dissipation rate comes mostly from interpolation because the through flow rate needed to balance the sedimentation of these particles does not match the prior experiments where the energy dissipation rate was measured. \citet{Kramel2018} provides a detailed description of the water tunnel used in these experiments and its working parameters.   

The particles used in the 3D experiments are also triads and are shown in Fig. \ref{triads} with $L=11$ mm and  $D=1.1$ mm which tapers down to a diameter of $0.7$ mm at the weak joints.
The particles were commercially printed by Rapid Prototype plus Manufacturing LLC (rp+m) using their TangoBlackPlus FLX980 polyjet material.
The particle density is approximately 1.12-1.13 g/cm$^3$ so they settle slowly in water.  These triads have a torsion coefficient of $k=1.3 \times 10^{-5} \;\textrm{Nm}/\textrm{rad}$ and are made from a polymer with Shore durometer of 26A which converts to a Young's modulus of 0.97 MPa using the relationship derived by \citet{Qi2003}.
The dimensions of the triads used are limited by the 3D printer resolution and are not as small as the Kolmogorov length scale of our flow ($\eta=0.28 \pm 0.05$ mm), and as a result they could only measure coarse-grained velocity gradients.

We track the particles using four high speed Phantom VEO 640S cameras and backlight illumination. Each camera has 72 Gb of RAM and records 1400 frames per second at a resolution of $1600 \times 2560$ pixels, and so a camera records a total of 12470 frames at full resolution until the internal RAM is filled. 
After recording, the cameras transfer the recorded data to two data servers for later analysis. 

\subsection{3D Results}
\label{3D results}

\begin{figure*}
\centering
 \includegraphics[width=0.85\textwidth]{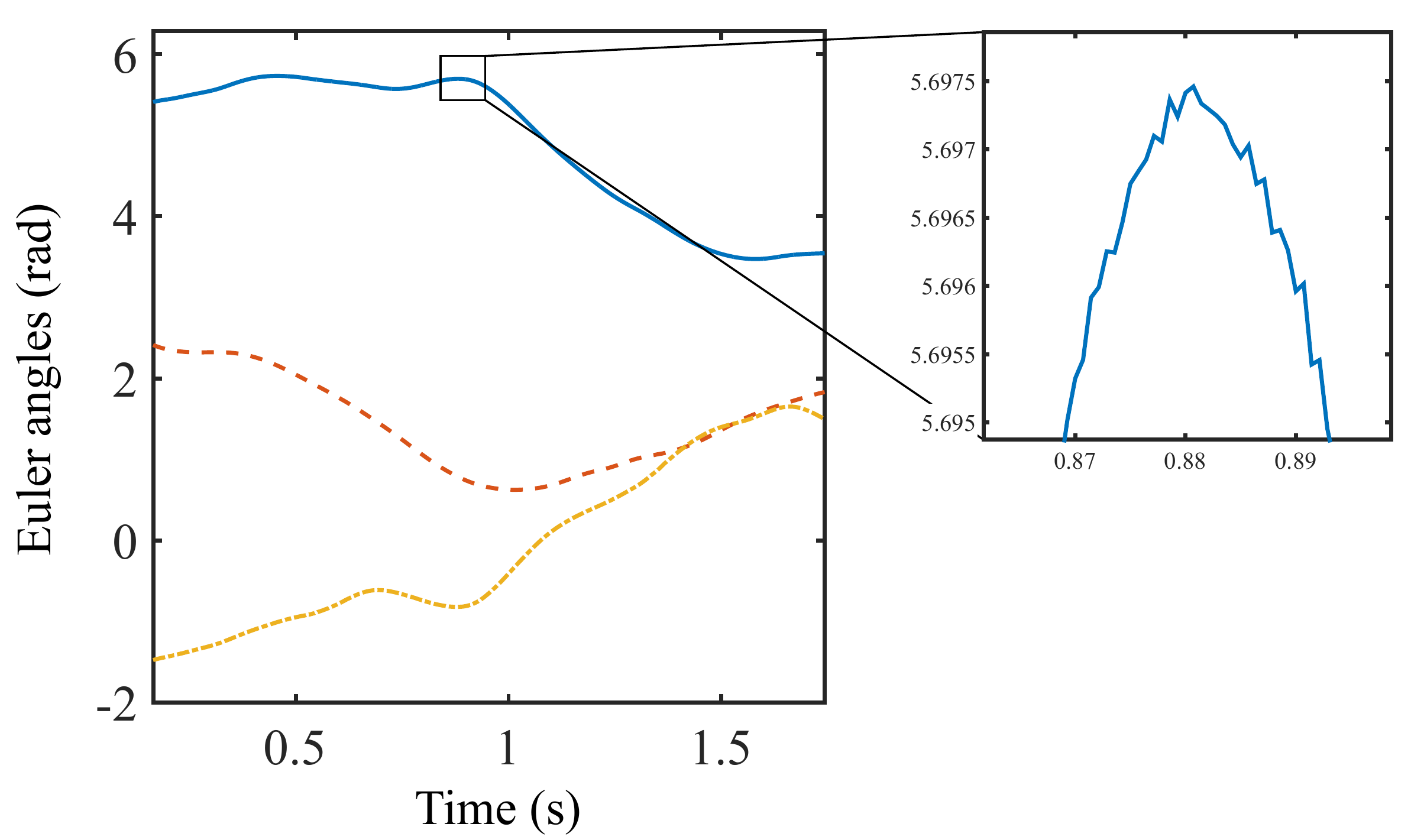}
\caption{The Euler angles giving the particle orientation as a function of time. The expanded view shows the accuracy in obtaining the particle orientations to be on the order of $10^{-4}$ radians.}
\label{Euler angles}
\end{figure*}

Using our model of Eq.~\ref{torque balance} we can calculate expected arm deformations for the experiments done in 3D turbulence.
We can estimate the strain rate of the flow from the energy dissipation rate using $\epsilon = 15\nu\langle(\partial u/ \partial x)^2\rangle$ ~\citep{Pope2000}. 
With these parameters, we calculate that an arm of a triad would deform by approximately $4.5 \times 10^{-4}$ radians.

In order to measure the extremely small deformations experienced by a triad in our flow we require a high level of accuracy in determining the position and orientation of a particle.
Fig.~\ref{Euler angles} shows the three Euler angles that determine the orientation of a triad as a function of time. The expanded view of Fig.~\ref{Euler angles} shows the random uncertainty in particle orientations to be on the order of  $10^{-4}$ radians, which is within the same order of magnitude as our predicted deformations.
We determine the position and orientation of a triad by using a least squares fitting routine of a modeled triad in three dimensions to the four images of the triad.
This high level of accuracy in measuring particle orientations is achieved by creating a model triad that closely resembles the actual triad used in the experiments.

\begin{figure*}
\centering
 \includegraphics[width=0.99\textwidth]{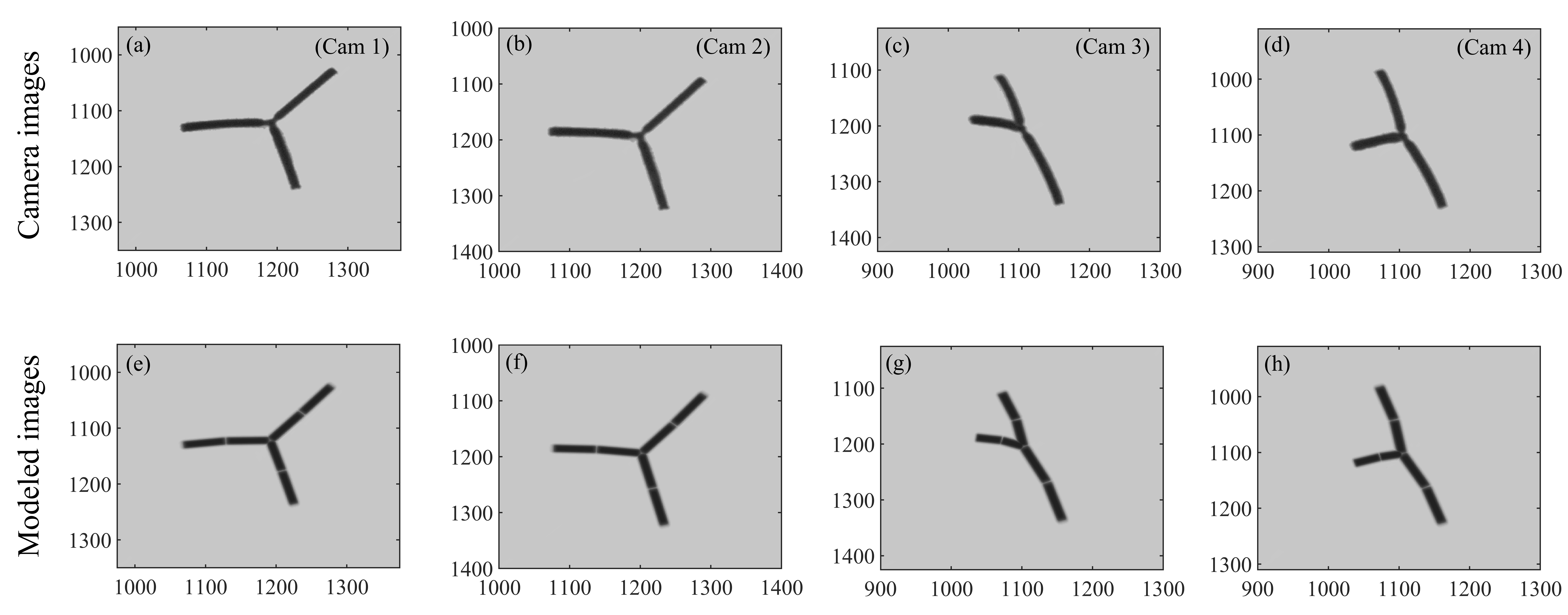}
\caption{Top row: the background subtracted image of a particle in a single frame as seen on all four cameras. Bottom row: the modeled particle fitted to the image at the same instance of time with two segmented arms to capture the natural curvature that exists in the arms of the particle.}
\label{model images}
\end{figure*}

Figure~\ref{model images}(a-d) show experimental images of a particle in one frame as seen by all four cameras, with cropping and background subtraction. As we see from the images, the triads used in these experiments are not ideal. The triads are made of soft polymeric materials that naturally have built in strains and imperfections which cause each arm of a particle to have a distinct curvature and shape.  Deformation by the fluid still occurs at the weak link, but each arm has a constant equilibrium shape that is not a straight rod.  We incorporate the unique curvature and shape of each arm at equilibrium into our model by creating arms that are made up of two separate segments.
The overall orientation and shape of each arm is given by 4 angles, 2 of which are associated with the rotation of the whole arm about the base of the particle and 2 are rotations of the second segment of the arm which give an arm its distinct curvature and shape. 
For this proof of concept experiment, we determined the equilibrium, non-deformed angles for one particle manually.  We chose a particle that remained in view for 1.7 seconds and averaged the orientation and shape of each arm through frames selected from the full duration that the particle remained in the detection volume. We then fix the angles that determine the unique shape of a particle in its non-deformed state so we can then accurately measure deformations that occur about this equilibrium state.
Fig.~\ref{model images}(e-h) show the model particle fitted to the image above it as seen by all four cameras.
We could further increase the accuracy of our model by increasing the number of segments that build up an individual arm and thus create a more realistic model, but as is evident by the data and Fig.~\ref{model images}, a two segment arm adequately captures the curvature in an arm. 

\begin{figure}
\centering
 \includegraphics[width=0.4\textwidth]{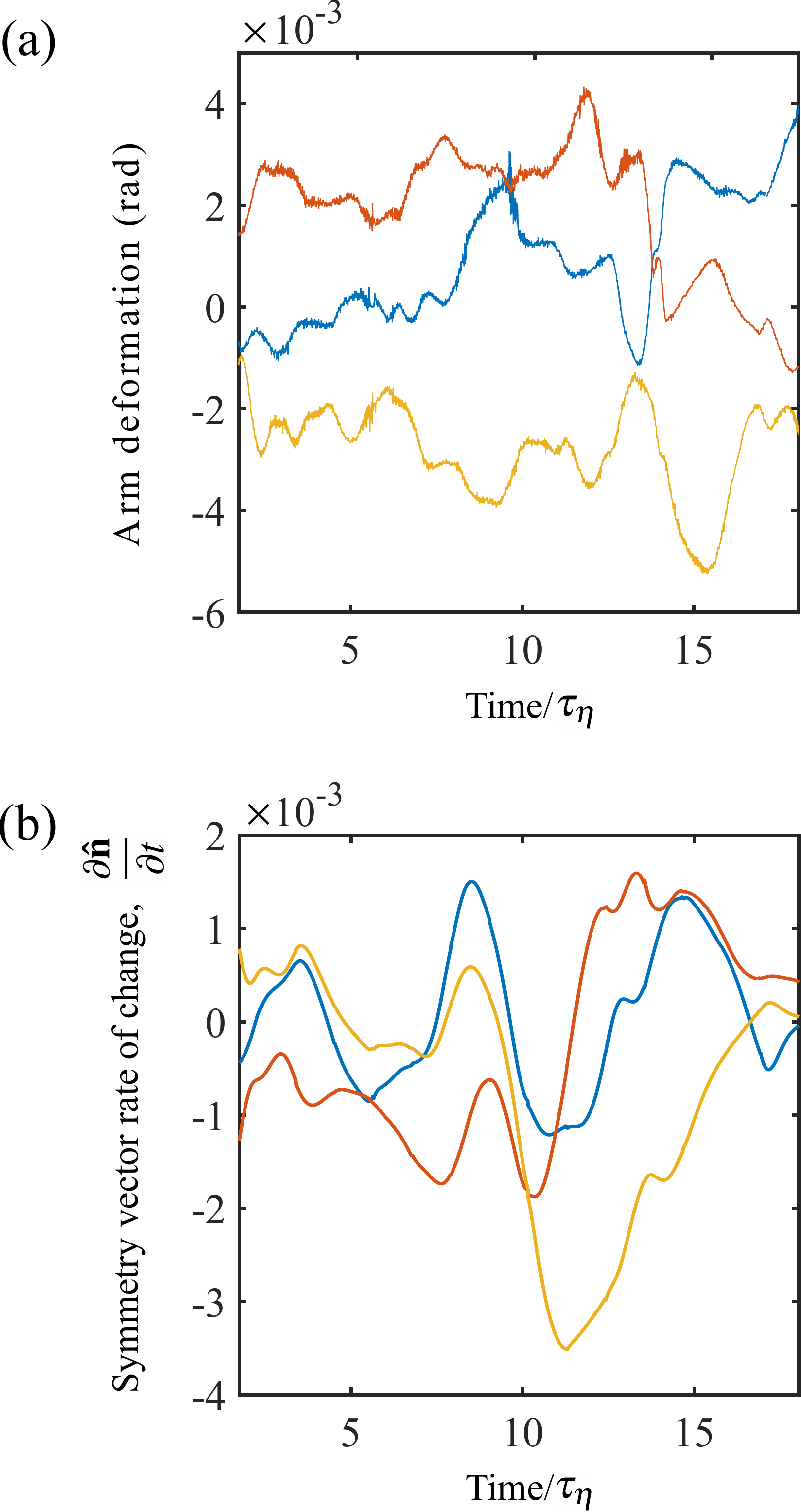}
\caption{(a) Measured deformations for the three arms of a triad as a function of time. Each arm is deformed by 0.001-0.002 radians. (b) $x$, $y$, $z$ components of $\partial \mathbf{\hat{n}}/\partial t$ plotted as a function of time for the triad in part (a). $\mathbf{\hat{n}}$ is the unit vector along the symmetry axis perpendicular to the plane of the triad.}
\label{3Dexp}
\end{figure}

Figure~\ref{3Dexp}(a) shows the measured deformations for each arm of a triad as a function of time. We can see that each arm is deformed by 0.001-0.002 radians.  
The measured deformations are close to our predicted value and the orientation measurement accuracy, making it difficult to know if the arms are actually deforming or if the observed deformations are due to noise in our particle orientation measurements. 
One way to qualitatively evaluate whether the deformations observed in Fig.~\ref{3Dexp}(a) are caused by turbulent strain or by measurement errors is to compare with the observed rotation rate of the particle.  In turbulence, both the vorticity and the strain rate change on the Kolmogorov time scale and so turbulence induced deformations should have a correlation time similar to turbulence induced rotations.  Fig.~\ref{3Dexp}(b) shows the time derivative of $\mathbf{\hat{n}}$ as a function of time, where $\mathbf{\hat{n}}$ is the symmetry axis perpendicular to the plane of the triad.   The variations in deformations and rotations have roughly similar time scales.  There is shorter time scales in the arm deformation, but a factor of two faster variation in arm deformation than vorticity is possible since strain is a second rank tensor while vorticity is a vector.  For example, Fig.~\ref{def trailing arm} shows that the deformation has two maxima in each rotation of a particle in simple shear while for typical tumbling motion in 3D turbulence, the components of $\frac{\partial \mathbf{\hat{n}}}{\partial t}$ have only one maximum per period.    So we conclude that the measured deformations shown in Fig.~\ref{3Dexp}(a) are likely dominated by real deformations caused by the turbulence.

\begin{figure}
\centering
 \includegraphics[width=0.4\textwidth]{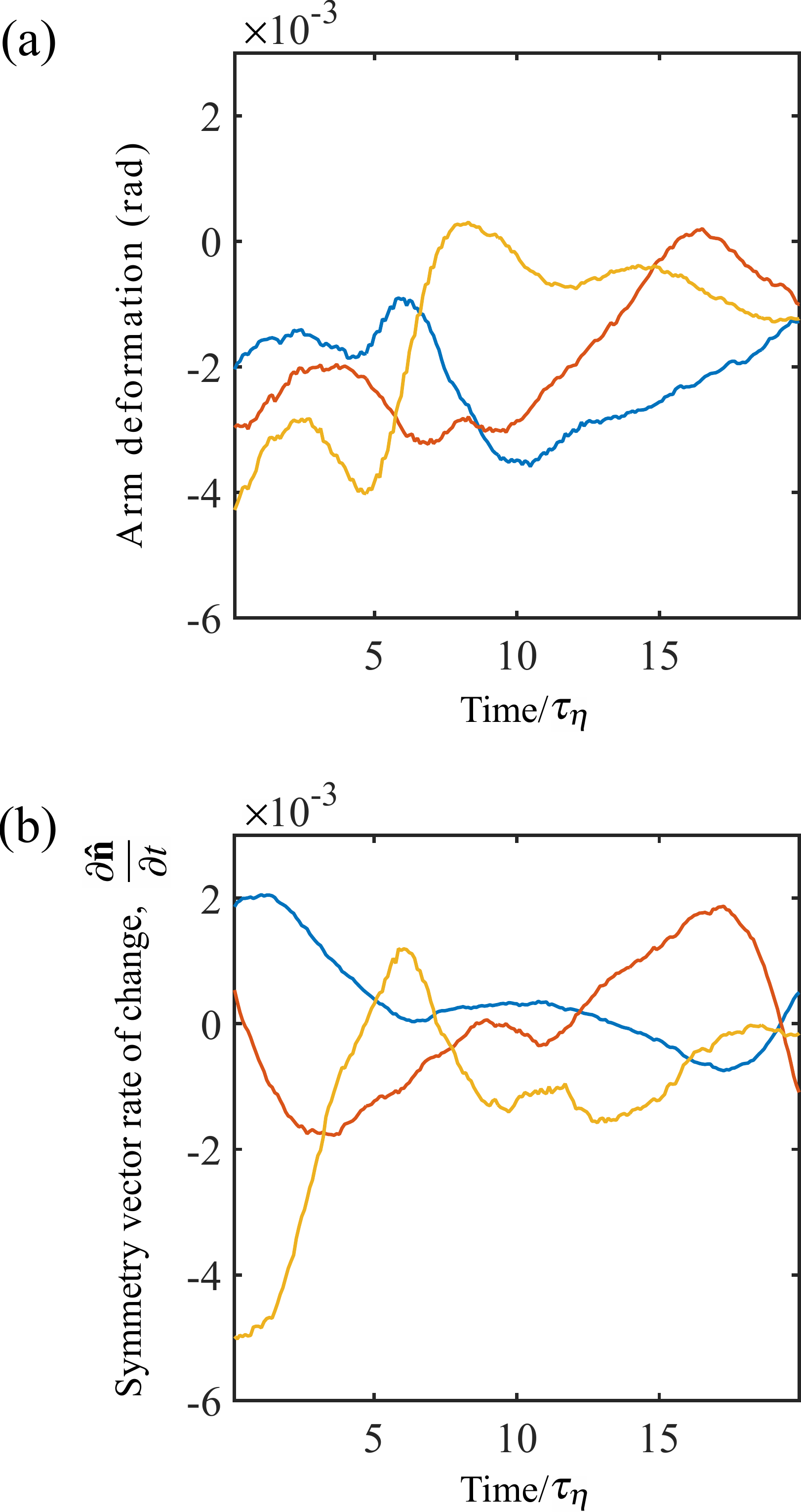}
\caption{(a) Measured deformations for the three arms of a simulated triad in turbulent flow. (b) $x$, $y$, $z$ components of $\partial \mathbf{\hat{n}}/\partial t$ plotted as a function of time for the simulated triad in (a).}
\label{3Dsim}
\end{figure}

Figure~\ref{3Dsim}(a) shows the deformation of the arms of a triad simulated in turbulence from the Johns Hopkins turbulence database~\citep{Li2008,Yu2012}.
We simulate the motion of a deformable triad in forced isotropic turbulence on a $1024^3$ triply periodic box.  The particles are modeled as following a Lagrangian trajectory and rotating and deformaing in response to the velocity gradients along their trajectories. 
The overall orientation of a triad as it is advected by the flow can be determined using the left Cauchy-Green strain tensor $\mathbf{C}^{(L)}$, which is calculated using the deformation gradient tensor $\mathbf{F}$~\citep{Ni2014}.
The eigenvector corresponding to the maximum eigenvalue of $\mathbf{C}^{(L)}$ aligns with the most extensional direction of the fluid stretching, and fibers tend to align themselves with this direction. 
However, the effective ellipsoid of a triad is a disk and the unit vector along the symmetry axis of a disk aligns with the eigenvector that corresponds to the minimum eigenvalue of $\mathbf{C}^{(L)}$.
Thus by calculating the eigenvectors of $\mathbf{C}^{(L)}$ we obtain the orientation of the symmetry axis of a triad at every instance along its trajectory.

We can use $\mathbf{F}$ and $\mathbf{C}^{(L)}$ to solve for the rotation tensor $\mathbf{R}$ which by acting on the triad will account for the rotation of the triad arms in the plane of the triad.
With the orientation of the particle and the position of its arms determined, we can now calculate the arm deformations due to fluid strain. We normalize the simulation strains using the eddy turnover time at the scale of the particle $\tau_{L}$ to match our experimental strain values.
Since a triad is not an isotropic particle and does not rotate with the same rotation rate as the fluid vorticity, we use the general expression for arm deformations from Eq.~\ref{torque balance}.

Comparing the simulated deformations in Fig.~\ref{3Dsim} with the exprimental deformations in Fig.~\ref{3Dexp}, we see that both have the same order of magnitude suggesting that the deformations in Fig~\ref{3Dexp}(a) are actual deformation and not measurement noise.
The experimental deformations have a somewhat faster time scale which could be simply a feature of this trajectory or it could suggest that some of the measured deformations are non-physical.

The noise that is visible in the simulated data in Fig.~\ref{3Dsim} is due to interpolation difficulties in the database.  In those simulations of a $1024^3$ periodic box at $R_{\lambda}=418$, the grid spacing is 2.19$\eta$ and so there are often cases where a component of the velocity gradient tensor at neighboring grid points differ by 50\% of the rms velocity gradient component.  When these velocity gradient fields are interpolated onto Lagrangian trajectories, there are interpolation artifacts of a few percent.

While the 3D turbulence experiments provide evidence of the high level of accuracy that can be achieved in measuring particle position and orientation, we do not observe substantial deformations with the current parameters. Our techniques for measuring velocity gradients from particle deformations are promising, and future work should be able to study more flexible particles in larger strain rates where accurate determination of 3D velocity gradient tensors would be possible.

\section{Conclusions}
\label{conclusions}

Deformable particles exhibit unique responses to the strain and vorticity components of uniform velocity gradient tensors, allowing for the reconstruction of the full velocity gradient tensor from a single particle.  Using a triad particle in a 2D simple shear flow, we were able to accurately measure arm deformations and extract the instantaneous velocity gradient tensor from the measurements. 
Using smaller triad particles in 3D turbulence behind an active jet array, we obtained highly accurate measurements of particle arm orientations on the order of $10^{-4}$ radians and evidence that we are able to measure arm deformations by the turbulence.  We compared our experimental results with simulations of deformable triads which showed similar deformations.

While these measurements are not yet able to measure statistics of particle deformations and velocity gradients in turbulence due to the calibration process required for individual particles, we have demonstrated the potential of this method for future measurements.   We anticipate that future work will be able to obtain more flexible particles in the shape of tetrads whose manufacturing variations are minimal.  Measurement of such particles in a flow with larger velocity gradients should allow a powerful new way to measure 3D velocity gradient tensors from single particle measurements.

\section{Appendix}
\label{appendix}
To model the deformation of an arm with a weak link, consider a weak link formed by a short cylindrical segment of the arm of length $L_J$ with $L_J \ll L$ and diameter $D_J$ with $D_J=cD$ and $c < 1$.   Our weak links are actually made with more complex shapes as shown in Fig.~\ref{triads} that are simpler to 3D print, but this model is adequate for the scaling argument.
If there is negligible bending away from the weak link, the spring constant is determined only by the joint, $k \propto EI_{J}/L_{J}\propto ED_{J}^4/L_{J}$.
Here $I_{J}$ is the second moment of area of the joint.
Using Eq.~\ref{deltheta} and substituting for $k$ we have 
\begin{equation}
\delta \theta\propto \frac{\mu}{E\tau_{\eta}} \frac{L_J}{c^3 D_J} \frac{L^3}{D^3}.
\end{equation}
So the arm deformation angle is proportional to the third power of the aspect ratio of the arm with a constant of proportionality, $L_J/(c^3D_J)$ determined by the geometry of the weak link.

For these arms with weak links, the buckling transition is different than for uniform cylinders.  However, to demonstrate that we are far from the buckling transition, we simply calculate the diameter of a uniform arm that would have the same spring constant and use the effective diameter to determine the value of $Z$ for our arms.   Since our values of $Z$ are more than four orders of magnitude smaller than typical buckling transitions, we know that buckling is not relevant in this case.

\begin{acknowledgements}
This work was supported by NSF grant DMR-1508575, and Army Research Office grants W911NF-15-1-0205 and W911NF-17-1-0176. We benefited from discussions with Stefan Kramel and Lee Walsh on software development and data analysis.
\end{acknowledgements}

\bibliographystyle{spbasic}      

\end{document}